\documentclass{emulateapj}
\usepackage{graphicx}

\def\be{\begin{equation}}
\def\ee{\end{equation}}
\def\ba{\begin{eqnarray}}
\def\ea{\end{eqnarray}}
\def\go{\mathrel{\raise.3ex\hbox{$>$}\mkern-14mu
             \lower0.6ex\hbox{$\sim$}}}
\def\lo{\mathrel{\raise.3ex\hbox{$<$}\mkern-14mu
             \lower0.6ex\hbox{$\sim$}}}
\def\etal{et al.\ \rm}

\begin{document}

\title{Structure of Passive Circumstellar Disks: Beyond the 
Two-Temperature Approximation.}
\author{Roman R. Rafikov\altaffilmark{1,2,3} and Fabio De Colle\altaffilmark{4,5}}
\altaffiltext{1}{CITA, McLennan Physics Labs, 60 St. George St., University of Toronto, Toronto, ON M5S 3H8, Canada; rrr@cita.utoronto.ca}
\altaffiltext{2}{IAS, Einstein Dr., Princeton, NJ 08540 USA}
\altaffiltext{3}{Canada Research Chair}
\altaffiltext{4}{Dublin Institute for Advanced Studies, 5 Merrion Square Dublin 2, Ireland; fdc@cp.dias.ie}
\altaffiltext{5}{Instituto de Astronom\'\i a, 
Universidad Nacional Autonoma de Mexico, Ap.P. 70264, 04510 DF, M\'exico}


\begin{abstract}
Structure and spectral energy 
distributions (SEDs) of externally irradiated circumstellar
disks are often computed on the basis of the two-temperature
model of Chiang \& Goldreich. We refine these 
calculations by using a more realistic temperature 
profile which is continuous at all optical depths and thus 
goes beyond the two-temperature model. 
It is based on the approximate solution of the radiation 
transfer in the disk obtained from the frequency-integrated 
moment equations in the Eddington approximation.  
We come up with a simple procedure (``constant $g_z$ approximation'') 
for treating the vertical structure of the disk in regions 
where its optical depth to stellar radiation   
is high. This allows us to obtain expressions for 
the vertical profiles of  density and pressure at every point 
in the disk and to determine the shape of its surface. 
Armed with these analytical results we calculate the full 
radial structure of the disk and demonstrate that it 
favorably agrees with the results of direct numerical calculations. 
We also describe a simple and efficient way of the SED calculation 
based on our adopted temperature profile. 
Resulting spectra provide very good match (especially at short 
wavelengths) to the results of more detailed (but also more time-consuming) 
SED calculations solving the full frequency- and 
angle-dependent radiation transfer within the disk.
\end{abstract}



\section{Introduction}
\label{sect:intro}


Protoplanetary disks around T Tauri stars (stellar mass 
$M_\star\lesssim 2~M_\odot$) as well as the disks 
around more massive Herbig Ae/Be stars ($M_\star\gtrsim 
2~M_\odot$) belong to the class of circumstellar disks that are 
strongly affected by the radiation of 
their central stars. Irradiation modifies both the vertical 
structure of the disk and the radial dependence of 
disk properties. As a result, observational 
signatures of centrally irradiated discs are quite different 
from those of the conventional viscously heated 
$\alpha$-disks (Shakura \& Sunyaev 1972).

Kenyon \& Hartmann (1987) have pointed out that the observed infrared 
spectra of protoplanetary disks can be understood as resulting from 
passive reprocessing of the stellar radiation assuming that 
disk surfaces have flared geometry. 
Later Chiang \& Goldreich (1997; hereafter CG97) have analytically 
derived hydrostatic, radiative equilibrium models of disks around 
T Tauri stars.  Their calculation was based on a {\it two-temperature 
approximation} for the disk thermal structure in which the outer
``superheated'' layer has been assumed isothermal at the temperature 
characteristic for dust grains directly illuminated by 
unattenuated stellar radiation. The isothermal midplane region 
was found to have lower temperature set by the balance between 
the oblique stellar illumination of the disk surface and the isotropic 
infrared emission of the disk interior. The success of this 
simple model has led to its wide acceptance for interpreting the  
observations of circumstellar disks.

The goal of this paper is to improve our understanding of 
the circumstellar disk structure by going beyond the simple 
two-temperature approximation of CG97. Calvet \etal (1991; 
hereafter C91) have studied the vertical radiative transfer 
within irradiated dusty circumstellar disks and obtained
approximate analytical expression for the vertical temperature
profile which is superior to the two-temperature approximation of
CG97. In this work we employ this temperature 
distribution, which is reviewed in detail in \S 
\ref{subsect:rad_transfer}, to calculate the disk structure
and spectrum. Under certain conditions it is possible to
derive approximate analytical solutions for the vertical
disk structure as demonstrated in \S \ref{sect:hydro_eq}. 
This allows us to efficiently compute the radial 
dependence of various disk properties in \S 
\ref{sect:pure_radial} and to compare them with purely
numerical solutions. Finally, in \S \ref{sect:spectra}
we demonstrate how the use of the realistic temperature 
profile of C91 improves the calculation of 
the spectral energy distribution (SED) of circumstellar 
disks.

Throughout this study we assume the major heating 
source of the disk to be the radiation of the central star
of mass $M_\star$, having radius $R_\star$ and effective 
temperature $T_\star$. Viscous heating is considered 
subdominant. Circumstellar disk extending from $R_{in}$
to $R_{out}$ is assumed to be 
geometrically thin thus allowing us to neglect radial 
transport of radiation. This simplifies radiative 
transfer within the disk and reduces it to a 
one-dimensional problem. Surface density of the disk $\Sigma_0$ 
varies as
\ba
\Sigma_0(a)=\Sigma_\star\left(\frac{R_\star}{a}\right)^{\delta},
\label{eq:Sigma}
\ea
where $a$ is the distance from the central star and 
$\Sigma_\star\equiv \Sigma(R_\star)$. We use $\delta=1$ in 
our calculations. 
Our results will be illustrated using three fiducial models 
of circumstellar disks around Herbig Ae/Be star, T Tauri star, 
and a brown dwarf, analogous to those adopted by 
Dullemond \& Natta (2003; hereafter DN03).
Properties of these models are summarized in Table 
\ref{table} 
($\Sigma_1\equiv\Sigma_\star(R_\star/1~\mbox{AU})^{\delta}$ is
the value of $\Sigma_0$ at $1$ AU).


\section{Thermal structure.}
\label{subsect:rad_transfer}


To describe the transfer of stellar radiation 
with blackbody temperature $T_\star$ within the disk we 
introduce optical depth $\tau_\star(z)$ ($z$ is the 
altitude above the disk midplane) which is 
calculated along the {\it radial} direction from the star. Following
conventional wisdom we approximate $\tau_\star$ as 
\ba
\tau_\star(z)\approx \alpha^{-1}\int\limits_z^\infty
\kappa_P(T_\star)\rho(z) dz = \alpha^{-1}\kappa_P(T_\star)\Sigma(z),
\label{eq:hor_vert_opt_depth}
\ea
where $\rho(z)$ 
is the local gas density, $\Sigma(z)\equiv \int_z^\infty
\rho(z) dz$ is the disk surface density above a given
$z$, $\kappa_P(T)$ is the Planck mean opacity at temperature $T$,
and $\alpha$ is the flaring angle ($\pi/2-\alpha$ is the 
angle between the normal to the disk surface and the radial 
direction). In the thin disk approximation
\ba
\alpha\approx \eta\frac{R_\star}{a}+a\frac{d}{da}
\left(\frac{H_1}{a}\right),
\label{eq:alpha}
\ea
where $H_1(a)$ is the altitude of the disk surface defined as 
a surface at which $\tau_\star(a,H_1)=1$ and $\eta\approx 0.4$ 
is constant (Kenyon \& Hartmann 1987; CG97). There are two regimes 
of irradiation: ``lamp-post'' illumination when 
$R_\star\gtrsim H_1$ and 
\ba
\alpha\approx \eta\frac{R_\star}{a},
\label{eq:lamp_post}
\ea
and ``point-source'' illumination when $R_\star\lesssim H_1$ and 
\ba
\alpha\approx a\frac{d}{da}
\left(\frac{H_1}{a}\right).
\label{eq:point_like}
\ea
In the first case the geometry of illumination is determined by 
the finite size of the source, while in the second it is 
the self-adjustment of the disk surface geometry that sets the 
illumination angle. Apparently, disk must be 
{\it flared} in the latter case, i.e. $d(H_1/a)/da>0$. 

\begin{figure}
\plotone{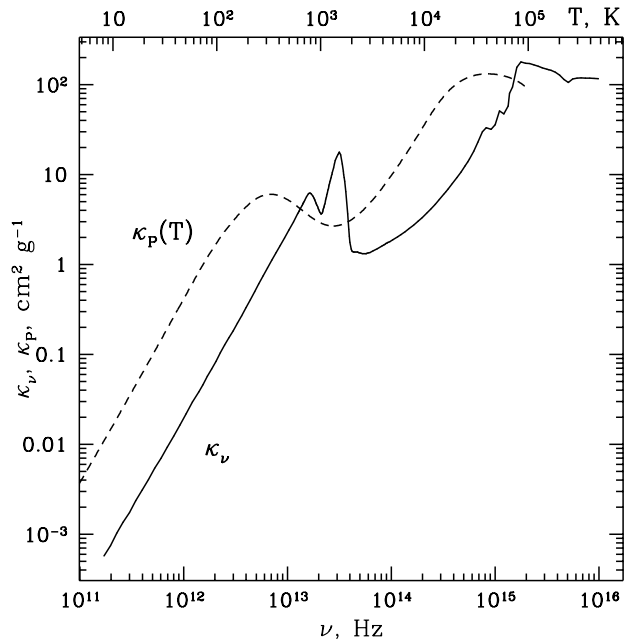}
\caption{
Opacity $\kappa_\nu$ ({\it solid curve}) adopted in this paper as a 
function of 
frequency $\nu$ ({\it lower axis}) and Planck opacity $\kappa_P$  
({\it dashed curve}) computed based on this $\kappa_\nu$ 
as a function of temperature $T=h\nu/k$ ({\it upper axis}). 
\label{fig:opacities}}
\end{figure}

Dust grains are the major opacity source in circumstellar
disks except their central regions where high temperature leads to 
dust sublimation. To compute $\kappa_P(T)$ we
use the frequency-dependent opacity $\kappa_\nu$
characteristic of $0.1$ $\mu$m silicate grains (Draine 
\& Lee 1984) with no scattering (the same as $\kappa_\nu$ 
used by DN03\footnote{See 
http://www.mpia-hd.mpg.de/homes/dullemon/radtrans/}, 
to facilitate subsequent comparison with 
their results). Both $\kappa_P(T)$ and $\kappa_\nu$ used in 
this study are shown in Figure \ref{fig:opacities}.

At low temperatures ($T\lesssim 10^2$ K) $\kappa_P$ behaves as a power 
law of $T$. Motivated by this, in deriving our analytical 
results we will frequently use the ``local power law opacity'' 
approximation to $\kappa_P(T)$. It is defined as the best 
power law fit to the actual run of $\kappa_P(T)$ in the 
temperature range relevant at a given distance from the 
star $a$:
\be
\kappa_P(T)\approx\kappa_0(a) T^{\beta(a)},
\label{eq:kappa_pow}
\ee
where parameters $\kappa_0$ and $\beta$ are functions of $a$. This 
simplified approach shows very good agreement with numerical 
calculations using full $\kappa_P(T)$, see \S 
\ref{sect:pure_radial}.

Calvet \etal (C91) have derived approximate 
temperature structure of centrally irradiated accretion
disk heated from inside by viscous dissipation. Their result 
is based on the assumption that disk is optically thick to 
its own radiation with opacity independent of temperature.
For simplicity, in this study we neglect radiation 
scattering by dust and viscous heating. At the same time, 
the assumption of a geometrically thin disk allows us to 
generalize C91 result for an {\it arbitrary} dependence 
of $\kappa_P$ on $T$. Simple calculation in the spirit of C91 results
in the following implicit expression for $T$:
\ba
T^4(z)=\phi T_\star^4\left(\frac{R_\star}{a}\right)^2\left[\frac{\alpha}{2}
\frac{\psi_{ex}}{\psi_{sh}}+\frac{q(T)}{4} e^{-\tau_\star(z)}\right].
\label{eq:tem_calvet}
\ea
Here $q(T)=\kappa_P(T_\star)/\kappa_P(T)$ is a ratio of opacities 
at temperatures $T_\star$ and $T$, factor $\phi$ represents the
fraction of the stellar surface visible from a given point of the
disk surface, while $\tau_\star(z)$ and $\alpha$
are given by equations (\ref{eq:hor_vert_opt_depth}) \& (\ref{eq:alpha}). 
If the inner edge of the disk is close to the stellar surface --
an assumption that we  adopt in this paper -- then $\phi\approx 1/2$. 
Following Dullemond, Dominik \& Natta (2001; hereafter DDN), 
we have introduced correction 
factors $\psi_{ex}\le 1$ and $\psi_{sh}\le 1$ to make this 
solution valid even when the disk is optically thin to the radiation 
of its outer layer and/or its interior 
[situations not allowed in the original optically thick 
solution of C91 which can be recovered from 
eq. (\ref{eq:tem_calvet}) by setting $\psi_{ex}=\psi_{sh}=1$]. 
We provide details of the calculation of $\psi$-factors in Appendix 
\ref{ap:psi_factors}. 

\begin{figure}
\plotone{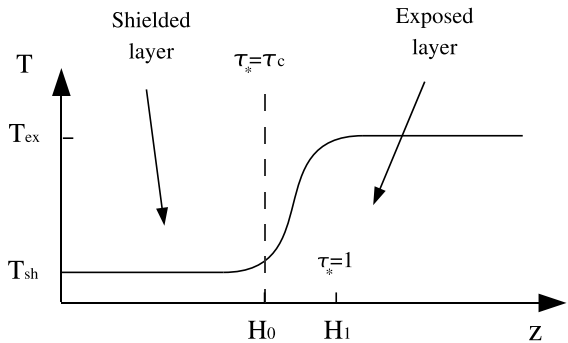}
\caption{
Schematic illustration of the vertical thermal structure 
of irradiated disk adopted in this paper. See text for detailed
explanations.
\label{fig:illustration}}
\end{figure}

According to equation (\ref{eq:tem_calvet}) the vertical extent
of the disk can be split into two layers: the outer layer 
{\it exposed} to stellar radiation and the inner layer 
{\it shielded} from direct starlight (see Figure \ref{fig:illustration}
for an illustration).

Dust within the {\it exposed} layer absorbs highly anisotropic 
stellar flux with blackbody temperature $T_\star$ and 
reradiates it isotropically at lower temperature. 
A half of the reradiated flux escapes 
while another half illuminates the interior of the disk.
Temperature in the exposed layer varies vertically according to
\ba
&& T\approx T_\star\left(\frac{R_\star}{2a}\right)^{1/2}
\left[\phi q(T) e^{-\tau_\star(z)}\right]^{1/4
\label{eq:T_trans}}\\
&& \approx T_{ex}(a)e^{-\tau_\star/(4+\beta)}.
\label{eq:T_trans_pow}
\ea
where
\ba
&& T_{ex}(a)=T_\star\left[\phi q(T_{ex})\right]^{1/4}
\left(\frac{R_\star}{2a}\right)^{1/2}
\label{eq:T_ex}\\
&& = T_\star\left(\frac{R_\star}{2a}\right)^{2/(4+\beta)}\phi^{1/(4+\beta)}.
\label{eq:T_ex_pow}
\ea
Expressions (\ref{eq:T_trans_pow}) 
and (\ref{eq:T_ex_pow}) correspond to the case of the 
power law opacity (\ref{eq:kappa_pow}). It follows from 
equation (\ref{eq:T_trans}) that $T_{ex}$ is the temperature 
of the optically thin part of the disk ($\tau_\star\ll 1$).
Exposed layer corresponds to $\tau_\star\lesssim 
\tau_c$ where
\ba
\tau_c\equiv\ln\left[\frac{q(T_{sh})}{2\alpha}
\frac{\psi_{sh}}{\psi_{ex}}\right].
\label{eq:tau_c}
\ea
We denote $H_0(a)
\equiv z(\tau_\star=\tau_c)$ the height at which optical
depth $\tau_\star$ equals $\tau_c$.

{\it Shielded} layer lies below the exposed region at high 
optical depth, $\tau_\star\gtrsim \tau_c$. 
Within this layer disk material receives 
only the reprocessed radiation of the exposed layer since the 
direct stellar flux is almost entirely absorbed by the
exposed layer. Shielded region is isothermal with temperature
\ba
T_{sh}(a)=T_\star\left(\frac{\phi\alpha}{2}
\frac{\psi_{ex}}{\psi_{sh}}\right)^{1/4}
\left(\frac{R_\star}{a}\right)^{1/2}, ~~~\tau_\star\gtrsim \tau_c. 
\label{eq:T_irr}
\ea
Equation (\ref{eq:T_irr}) results from balancing the
incoming flux of the reprocessed radiation with the energy loss 
from the shielded layer.
Factors $\psi_{sh}$ and $\psi_{ex}$ in equations (\ref{eq:tau_c})
and (\ref{eq:T_irr}) accounting for the possibility 
of optically thin disk are set by $T_{sh}$ and $T_{ex}$, 
see equations (\ref{eq:psi_irr_DDN}) and (\ref{eq:psi_ex_RD}).

Temperature profile (\ref{eq:tem_calvet}) should
be compared with the temperature structure adopted
by CG97:
\ba
T &=& T_{ex},~~~~~\tau_\star<1,\nonumber\\
&=& T_{sh},~~~~~\tau_\star\ge 1.
\label{eq:T_CG}
\ea
Equation (\ref{eq:tem_calvet}) provides a better match for the real
thermal structure of the exposed layer as it does not feature an unphysical 
temperature jump at some optical depth.


\section{Vertical structure.}
\label{sect:hydro_eq}


Hydrostatic equilibrium in $z$-direction 
is described by
\ba
\frac{dP}{dz}=-\Omega^2 z \rho,
\label{eq:hydro}
\ea
where $P$ is the gas pressure and 
$\Omega\equiv (G M_\star/a^3)^{1/2}$ is the 
local angular frequency. Within the isothermal shielded layer
($z\lesssim H_0$) this equation yields 
\ba
\frac{\rho(a,z)}{\rho_0(a)}=\frac{P(a,z)}{P_0(a)}=
\exp\left[-\frac{z^2}{2h_{sh}^2(a)}\right],
\label{eq:rho_P}
\ea
where $h_{sh}\equiv \Omega^{-1}(k_BT_{sh}/\mu)^{1/2}$ is the isothermal
scaleheight of the disk in the shielded region, and $\rho_0$ and
$P_0$ are the midplane values of gas density and pressure. Because of the
exponential decay of $\rho$ with $z$ shielded region contains 
most of the surface density $\Sigma_0(a)$ of the disk 
(if $H_0\gtrsim h_{sh}$) which implies that 
\ba
\rho_0(a)\approx (2\pi)^{-1/2}\frac{\Sigma_0(a)}{h_{sh}},~~~
P_0=\frac{\rho_0}{\mu}k_BT_{sh}.
\label{eq:rho_through_Sigma}
\ea


\subsection{Constant $g_z$ approximation.}
\label{subsect:g_z}

Exposed region is not isothermal when $\tau_\star\sim 1$ but we can 
still understand its vertical structure by introducing the so-called
{\it constant $g_z$ approximation}. This approximation makes use of 
the fact that gas density in the disk decreases very 
rapidly with increasing $z$, see e.g. (\ref{eq:rho_P}). 
As a result, when considering the hydrostatic equilibrium, 
we can to zeroth order neglect the variation of the vertical 
acceleration $g_z=\Omega^2 z$ compared to the change of $\rho$ 
with $z$ in equation (\ref{eq:hydro}). 

In particular, for the determination of the disk structure near 
its surface  we can set $z\approx H_1$ which allows us to 
integrate equation (\ref{eq:hydro}) over $z$. We find  that 
$P(z)\approx \Omega^2 H_1 \Sigma(z)$ and using definition
(\ref{eq:hor_vert_opt_depth}) obtain the following relation 
between $P$ and 
$\tau_\star(z)$:
\ba
P(\tau_\star)\approx \alpha \frac{\Omega^2 H_1}{\kappa_P(T_\star)}
\tau_\star. 
\label{eq:P_vs_tau}
\ea 
Substituting this into equation (\ref{eq:hydro}) one finds
\ba
\frac{1}{\tau_\star}\frac{d\tau_\star}{dz}\approx -z\frac{\Omega^2 \mu}{k_B
T(\tau_\star)}.  
\label{eq:P_vs_tau1}
\ea 

In the shielded region below $H_0$ (we assume that $H_0\sim H_1$ which 
is verified below) $T(\tau_\star)\approx T_{sh}$
and equation (\ref{eq:P_vs_tau1}) yields
\ba
\frac{z^2}{2h_{sh}^2}\approx \ln\left[\frac{P_0\kappa(T_\star)}
{\alpha \Omega^2 H_1}\right]
-\ln\tau_\star.
\label{eq:sol}
\ea 
Integration constant is fixed in equation (\ref{eq:sol})
with the aid of equation (\ref{eq:P_vs_tau}) by noticing that in
the shielded region solution (\ref{eq:sol}) must reduce to (\ref{eq:rho_P}). 
Substituting $\tau_\star=\tau_c$ into equation (\ref{eq:sol}) one obtains 
the height of the boundary between the
exposed and shielded layers $H_0$:
\ba
&& H_0=\lambda h_{sh},~~~~\lambda\equiv\left(2\ln\Lambda\right)^{1/2},
\label{eq:lambda}\\
&& \Lambda\equiv\frac{P_0\kappa(T_\star)}{\alpha\tau_c \Omega^2 H_0}=
\frac{\Sigma_0\kappa(T_\star)}{\sqrt{2\pi}\alpha\tau_c\lambda}.
\label{eq:Lambda}
\ea 
Our use of $H_0$ instead of $H_1$ in the definition of 
$\Lambda$ improves the accuracy of (\ref{eq:Lambda}). 

Structure of the exposed region will be determined under the assumption 
of the ``local power law opacity'' approximation (\ref{eq:kappa_pow}) which
allows us to use equation (\ref{eq:T_trans_pow}).
When integrating equation (\ref{eq:P_vs_tau1}) for $z>H_0$ we can 
split the integral into two parts: $\int^{\tau_\star}=
\int^{\tau_c}+\int^{\tau_\star}_{\tau_c}$. In the first integral we set
$T(\tau_\star)=T_{sh}$ which reduces it to equation (\ref{eq:sol}),
while in the second we use $T(\tau_\star)$ given by equation 
(\ref{eq:T_trans_pow}). As a result, for $z\gtrsim H_0$ one finds
\ba
z^2\approx H_0^2
-2h_{ex}^2\left[\rm{Ei}\left(-\frac{\tau_\star}{4+\beta}\right)-
\rm{Ei}\left(-\frac{\tau_c}{4+\beta}\right)\right],
\label{eq:sol1}
\ea 
where $h_{ex}\equiv \Omega^{-1}(k_BT_{ex}/\mu)^{1/2}$ is the
disk scale height corresponding to temperature $T_{ex}$ and 
${\rm Ei}(x)\equiv -\int^\infty_{-x}t^{-1}e^{-t}dt$ is 
exponential integral (Gradshtein \& Ryzhik 2000).

Equation (\ref{eq:sol1}) allows us to determine the position of 
the disk surface:
\ba
&& H_1^2\approx H_0^2+\chi^2 h_{ex}^2,
\label{eq:H_1}\\
&& \chi^2\equiv
-2\left[\rm{Ei}\left(-\frac{1}{4+\beta}\right)-
\rm{Ei}\left(-\frac{\tau_c}{4+\beta}\right)\right]>0,
\label{eq:chi2}
\ea 
where for all practical purposes factor $\chi\sim 1$ can be 
considered almost constant -- although $\tau_c\gtrsim 1$ is a 
function of $a$, the corresponding
contribution in (\ref{eq:chi2}) is small compared to the leading term
$\rm{Ei}(-1/(4+\beta))$. One can see that $H_1\sim H_0$ unless
$\lambda\lesssim \chi T_{ex}/T_{sh}$ (which can only be realized
when the constant $g_z$ approximation breaks down, see \S 
\ref{subsect:const} and Figures \ref{fig:radial_1}-\ref{fig:radial_3}).
Asymptotically 
\ba
{\rm Ei}(-x) &\simeq & -\frac{e^{-x}}{x}\left[1+O(x^{-1})\right],~~~x\gg 1,
\nonumber\\
&\simeq & {\bf C}+\ln(x)+O(x),~~~~~~ x\ll 1,
\label{eq:Ei_as}
\ea
(Gradshtein \& Ryzhik 2000) where ${\bf C}\approx 0.5772$ is the Euler constant.
These limiting forms suggest that
\ba
z^2\approx H_0^2+2(4+\beta)h_{ex}^2 \frac{e^{-\tau_\star/(4+\beta)}}
{\tau_\star},
\label{eq:sol_ps}
\ea  
for $1\lesssim \tau_\star\lesssim \tau_c$\footnote{This assumes 
that $\tau_c\gg 1$ which
may be a questionable assumption in real disks.} ($H_0\lesssim 
z\lesssim H_1$) 
while 
\ba
&& \tau_\star(z)\approx (4+\beta)e^{-{\bf C}}
\exp\left(-\frac{z^2-H_0^2}{2h_{ex}^2}\right),
\label{eq:tau_fe}\\
&& \frac{P(z)}{P_0}\approx \frac{\rho(z)}{\rho_0}\nonumber\\
&& \approx 
\frac{(4+\beta)e^{-{\bf C}}}{\tau_c}
\exp\left(-\frac{z^2-H_0^2}{2h_{ex}^2}-\frac{H_0^2}{2h_{sh}^2}\right),
\label{eq:sol_fe}
\ea 
for $\tau_\star\lesssim 1$ ($z\gtrsim H_1$).  
The dependence (\ref{eq:tau_fe}) is somewhat unexpected given that
the exposed region above $H_1$ is virtually isothermal at temperature 
$T_{ex}$. Apparently, this is an artifact of our adopted constant $g_z$ 
approximation. At the same time $P(z)$ and $\rho(z)$ given by 
(\ref{eq:sol_fe}) are in accord with the isothermality of this part
of the disk which is to some extent coincidental. 

Equations (\ref{eq:lambda})-(\ref{eq:sol_fe}) fully determine the 
vertical hydrostatic structure of the disk as well as 
the positions of $H_0(a)$ and $H_1(a)$ 
through the parameters of the disk and the star.


\subsection{Comparison with the two-temperature approximation.}
\label{subsect:compare}

It is instructive to compare the results of the previous
section with the disk structure according to CG97.
Their model exhibits discontinuous jump of temperature
at $\tau_\star=1$ and assumes upper disk layer ($\tau_\star<1$) to
be fully isothermal, see equation (\ref{eq:T_CG}). This layer
(``superheated dust layer'' in their nomenclature) plays the 
role of our exposed region, while the underlying region ($\tau_\star>1$)
should be identified with our shielded layer. In both layers
vertical profile of pressure is Gaussian and $\rho$ is 
discontinuous across the boundary defined by $\tau_\star=1$ 
(while $P$ is continuous there).
Apparently, $H_0=H_1$ in the CG97 model; applying constant 
$g_z$ approximation with their temperature structure (\ref{eq:T_CG}) 
one finds $H_1$ to be given by equation (\ref{eq:lambda}) 
but with $\Lambda$ replaced by
\ba
\Lambda_{\rm CG}=
\frac{\Sigma_0\kappa(T_\star)}{\sqrt{2\pi}\alpha\lambda_{\rm CG}}.
\label{eq:Lambda_CG}
\ea
The position of the disk surface computed in the two-temperature 
approximation deviates from that given by equation (\ref{eq:H_1}).
Difference is not very significant in the inner disk 
where $\lambda^2 h_{sh}^2\gg \chi^2 h_{ex}^2$. However,
at intermediate\footnote{At large radii neither CG97 nor 
constant $g_z$ approximations work well, see 
\S \ref{subsect:const}.} radii this is no longer valid 
and the two-temperature model underestimates $H_1$ 
compared to our more accurate expression (\ref{eq:H_1}).


\section{Radial properties of irradiated disks.}
\label{sect:pure_radial}


Based on analytical results obtained in the previous
section we can provide a complete description of the 
irradiated disk structure. This 
requires the determination of $T_{sh}(a)$ 
which can be done by substituting $H_1(a)$ given
by the expression (\ref{eq:H_1}) into (\ref{eq:alpha}), 
plugging $\alpha$ into (\ref{eq:T_irr}) and solving the resultant
differential equation for $T_{sh}(a)$. 

\begin{figure}
\plotone{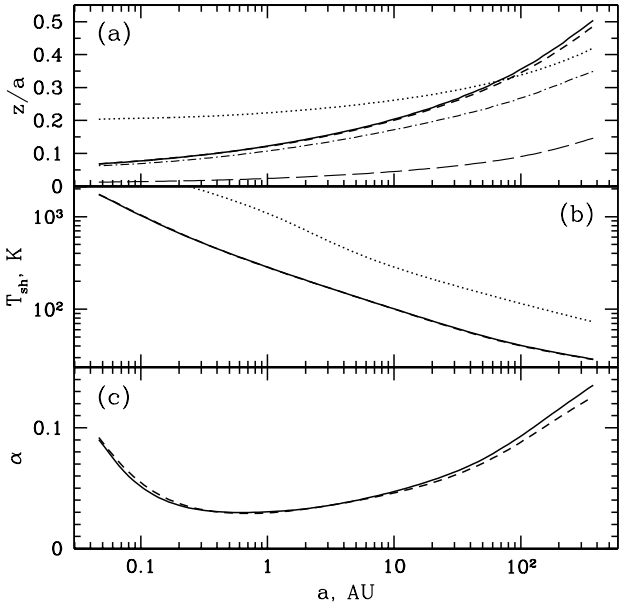}
\caption{
Comparison between the semi-analytical (constant $g_z$ approximation) 
and numerical approaches to computing the disk structure in Model I. 
(a) Position of the disk surface $H_1(a)/a$ calculated using 
eq. (\ref{eq:H_1}) ({\it solid curve}) and numerically
with the full treatment of the vertical hydrostatic equilibrium 
({\it short-dashed curve}). Also shown are the semi-analytical results 
for $H_0(a)/a$ ({\it dot-dashed line}), $h_{sh}(a)/a$ 
({\it long-dashed line}), and the ratio $h_{sh}/H_0=\lambda^{-1}$ 
({\it dotted line}).
(b) Midplane temperature $T_{sh}(a)$ computed semi-analytically 
({\it solid curve}) and numerically ({\it short-dashed curve}). Two methods 
give virtually identical results in this case. Dotted line represents 
$T_{ex}(a)$.
(c) Flaring angle $\alpha(a)$ obtained semi-analytically 
({\it solid curve}) and numerically ({\it short-dashed curve}).    
\label{fig:radial_1}}
\end{figure}

Analytical expressions for $T_{sh}(a)$ can be obtained 
for optically thick disks ($\psi_{ex}=\psi_{sh}=1$) 
which either (1) are illuminated in the lamp-post regime or
(2) experience point-like illumination and have 
$\lambda=H_0/h_{sh}$ of the order of several 
(typically $\gtrsim 3$). In the latter case one can neglect in equation 
(\ref{eq:H_1}) term $\chi h_{ex}$ compared to $\lambda h_{sh}$,
which, coupled with the weak sensitivity of $\lambda$ to $a$, leads 
to $H_1\propto h_{sh}$. This assumption should be valid in the 
inner, dense parts of circumstellar disks.
Analytical results for these two cases are summarized in 
Appendix \ref{ap:ap1}. In particular, one finds that 
$H_1\propto a^{9/8}$ for lamp-post illumination and $H_1\propto a^{9/7}$
for point-like illumination, i.e. in both cases disk surface is
flared. In a more general case we find the behavior of $T_{sh}(a)$ by 
solving the system of equations (\ref{eq:alpha}), 
(\ref{eq:T_irr}), and (\ref{eq:H_1}) numerically. 

To test the validity of the constant $g_z$ approximation we 
additionally compute disk structure without analytical 
approximations of \S \ref{subsect:g_z}, with the only 
assumption that the temperature profile is given by equation
(\ref{eq:tem_calvet}). In this approach at every radius 
we calculate vertical disk structure numerically 
with equations (\ref{eq:tem_calvet}) and (\ref{eq:hydro}), 
instead of using the results of \S \ref{subsect:g_z}. 
The value of $H_1$ obtained in this fashion 
(instead of using equation [\ref{eq:H_1}]) is then used
in equation (\ref{eq:alpha}) to reconstruct the radial 
disk profile. This approach uses 
the detailed opacity description $\kappa_P(T)$ shown
in Figure \ref{fig:opacities},  
while our semi-analytical procedure calculates disk structure 
assuming local power law opacity fit (\ref{eq:kappa_pow}) 
to the curve of $\kappa_P(T)$. Some details of the numerical 
disk structure calculation can be found 
in Appendix \ref{ap:ap3}. 

\begin{figure}
\plotone{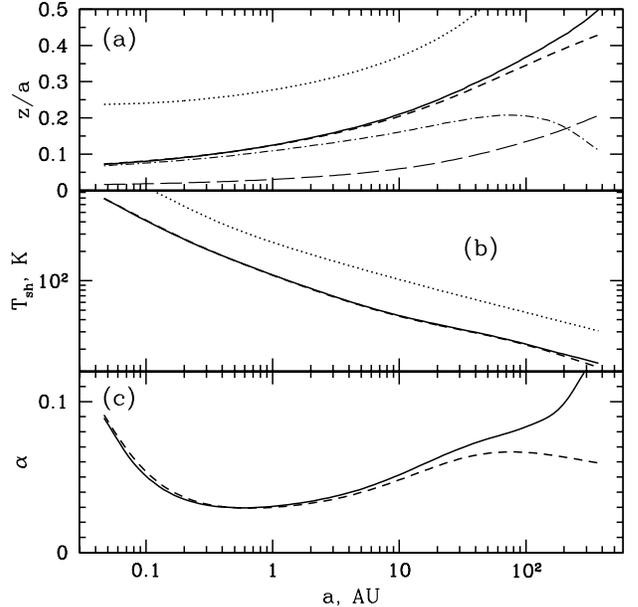}
\caption{
Same as Figure \ref{fig:radial_1} but for Model II.
\label{fig:radial_2}}
\end{figure}

In Figures \ref{fig:radial_1}-\ref{fig:radial_3} we present 
our results for the three disk models listed in Table 
\ref{table}. We compare the behaviors of $H_1(a)$, $T_{sh}(a)$,
and $\alpha(a)$ found semi-analytically and numerically. One can see 
that the theory based on the constant $g_z$ approximation 
predicts $H_1(a)$ quite 
well -- the discrepancy with the numerical result at the level of 
$\sim 10\%$ appears only in the outermost parts of the disk. The 
same is true for $T_{sh}(a)$. Flaring angle $\alpha$ is a 
more sensitive probe of the difference between the two approaches 
as it involves a 
derivative of $H_1(a)$. In the outer regions of disks in 
models II and III semi-analytical $\alpha(a)$ diverges significantly
from the numerical solution.  This is the result 
of the lower optical depth in these models compared to
the model I, which affects the validity of the constant $g_z$
approximation, see \S \ref{subsect:const}. 

\begin{figure}
\plotone{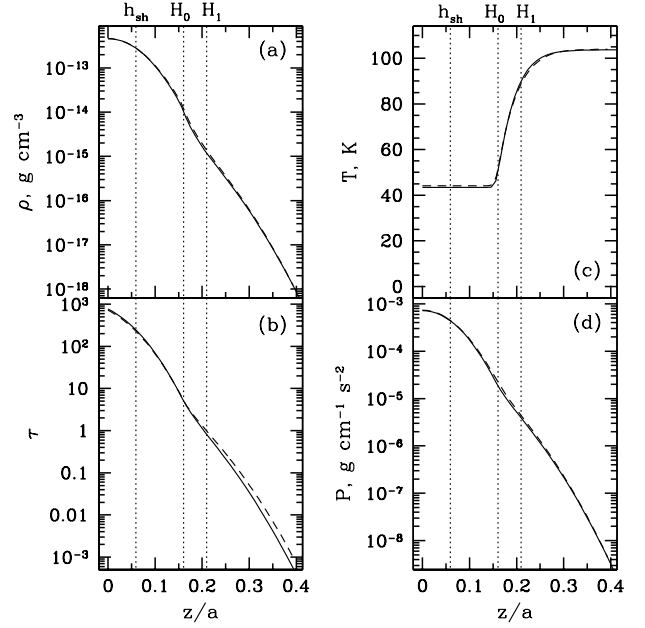}
\caption{
Vertical structure of the disk in Model II at $a=10$ AU. 
Results for $\rho(z)$ ({\it a}), $\tau(z)$ ({\it b}), $T(z)$ 
({\it c}), and $P(z)$ ({\it d}) are shown. Solid curves represent
fully numerical results while dashed curves represent results
obtained using constant $g_z$ approximation. Locations of 
$h_{sh}$, $H_0$, and $H_1$ are marked on each panel by dashed
lines. 
\label{fig:vertical}}
\end{figure}

Nevertheless, at small and intermediate $a$ the overall 
agreement between the two methods of the disk structure 
calculation is very good. For illustration, in Figure 
\ref{fig:vertical} we present the comparison between the 
vertical profiles of $T$, $P$, $\rho$, and $\tau$ obtained by 
both methods at $a=10$ AU in Model II. Despite the fact that 
at this location constant $g_z$ approximation is already close 
to the limit of its validity (see Figure \ref{fig:radial_2})
the agreement in the vertical structure of 
$P$, $\rho$, and $T$ is quite remarkable.
The constant $g_z$ approximation fails only for $\tau_\star(z)$
which was expected, see discussion after equation 
(\ref{eq:sol_fe}).


\subsection{Validity of the constant $g_z$ approximation.}
\label{subsect:const}

Constant $g_z$ approximation should work well when 
$H_0,H_1\gtrsim h_{sh}$ since in this case the decay of 
$P$ and $\rho$ with height is well represented by the 
tail of the Gaussian profile at $z\sim H_0,H_1$. As a result, 
at this height $\rho$ varies with $z$ 
much faster than $g_z$ and constant $g_z$ limit  
works well. Based on this reasoning, we can tentatively 
suggest $H_0\gtrsim 2h_{sh}$ as an approximate 
criterion for the validity of this approximation. 

\begin{figure}
\plotone{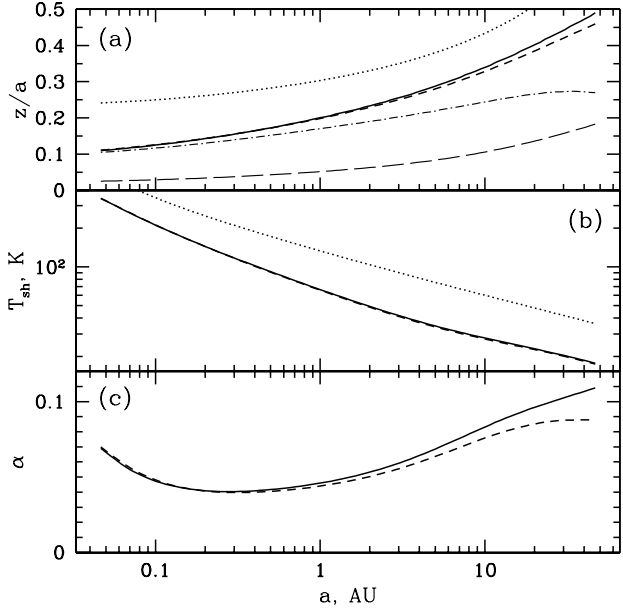}
\caption{
Same as Figure \ref{fig:radial_1} but for Model III.
\label{fig:radial_3}}
\end{figure}

This conclusion is confirmed by examination of Figures 
\ref{fig:radial_1}-\ref{fig:radial_3} which demonstrate 
that the relative difference between the values of 
$\alpha$ computed with and without constant $g_z$ approximation 
starts to exceed $10\%$ when the ratio $H_0/h_{sh}=\lambda$ 
drops below $\approx 2.5$. Using definition (\ref{eq:lambda}) 
we can then claim this approximation to be  
accurate (at the level of $\sim 10\%$) whenever 
$\Lambda\gtrsim 20$ or  
\ba
\kappa_P(T_\star)\Sigma_0\gtrsim 10,
\label{eq:validity}
\ea
where we have taken into account that in our models 
$\sqrt{2\pi}\alpha\tau_c\approx 0.5$ at the limit of
applicability. This condition is accurate only in the order of 
magnitude as the exact factor in its right hand side is 
exponentially sensitive to the uncertainties in $\lambda$. 
Despite this, in practical situations one may still use condition 
(\ref{eq:validity}) as a rough proxy for checking whether 
the constant $g_z$ approximation can be relied upon. 

\section{Spectral energy distribution.}
\label{sect:spectra}


\begin{figure}
\plotone{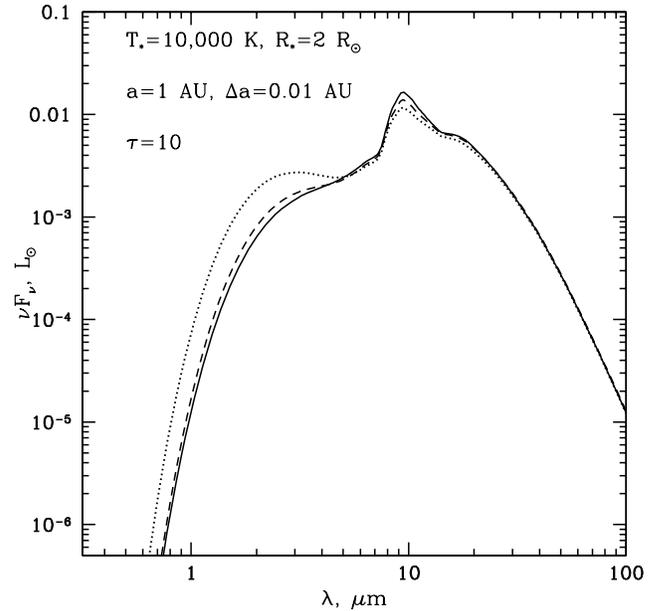}
\caption{
Spectral energy distribution of a disk annulus around
Herbig Ae/Be star ($T_\star=10^4$ K, $R_\star=2~R_\odot$,
$M_\star=2~M_\odot$).
Annulus is placed at $a=1$ AU, it has a width $\Delta a=0.01$ AU,
flaring angle $\alpha=0.05$,
and vertical optical depth $\tau=10$ at $550$ nm.  
Dashed line shows SED calculated according to equations 
(\ref{eq:SED_center}) and (\ref{eq:SED_outer1}),
solid line represents SED calculated using full frequency-
and angle-dependent radiation transfer (DN03), while dotted line 
is SED computed in the framework of the two-temperature 
approximation (CG97).
\label{fig:AeBe_1}}
\end{figure}

In this section we calculate disk
SED using our adopted temperature profile (\ref{eq:tem_calvet}), 
and then compare 
the outcome with the results of other approaches. 

In our case SED is produced by emission of both shielded and 
exposed layers. The 
contribution of the former per unit surface area of 
the disk $dF_\nu^{sh}(a)/dS$ is given simply by (see DDN)
\ba
\frac{dF_\nu^{sh}(a)}{dS}&=& 2\pi\cos i
\left\{1-\exp\left[-\frac{\Sigma_0(a)\kappa_\nu}{\cos i}\right]\right\}
\nonumber\\
&\times & B_\nu\left[T_{sh}(a)\right],
\label{eq:SED_center}
\ea
where $i$ is the disk inclination (we assume 
disk thickness to be so small that all elements of its 
surface can be characterized by a single value of inclination), 
$B_\nu(T)$ is a Planck 
function, and the factor in parentheses accounts for the 
possibility of the disk being optically thin at a given 
frequency $\nu$.

Exposed layer is always optically thin to its own radiation
and its contribution to the SED is given by 
\ba
\frac{dF_\nu^{ex}(a)}{dS}& = & 2\pi
\left\{1+\exp\left[-\frac{\Sigma_0(a)\kappa_\nu}{\cos i}\right]\right\}
\nonumber\\
&\times &\int\limits_0^{\tau_{\nu,c}} 
B_\nu\left[T(a,\tau_\star)\right]d\tau_\nu,
\label{eq:SED_outer}
\ea
where $\tau_\nu=\alpha\tau_\star\kappa_\nu/\kappa_P(T_\star)$ is
the optical depth at frequency $\nu$ along the disk normal, 
$T(a,\tau_\star)$ is given by (\ref{eq:T_ex}), and the factor 
in parentheses accounts for the contribution of the exposed layer 
on the other side of the disk when the shielded layer is optically thin 
at frequency $\nu$. 
Here $\tau_{\nu,c}=\alpha\tau_c\kappa_\nu/\kappa_P(T_\star)\ll 1$ 
is the value of $\tau_\nu$ at the boundary between the shielded 
and exposed layers.

\begin{figure}
\plotone{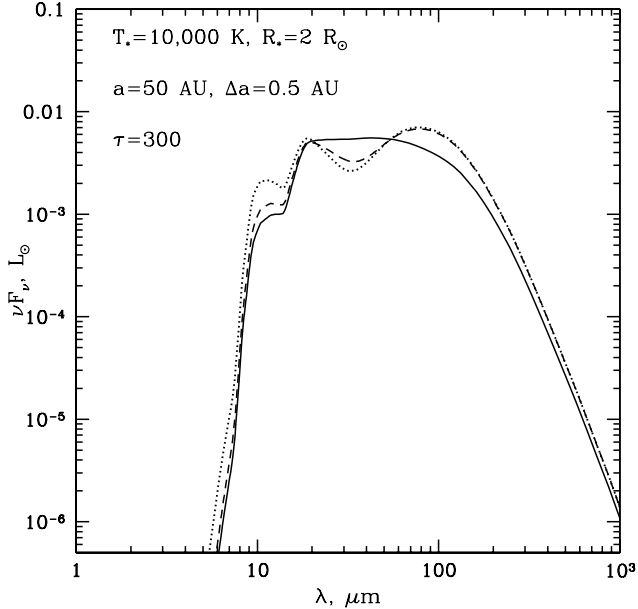}
\caption{
Same as Figure \ref{fig:AeBe_1} but for 
annulus at $a=50$ AU from the star having width 
$\Delta a=0.5$ AU and vertical optical depth $\tau=300$ at $550$ nm.  
\label{fig:AeBe_50}}
\end{figure}

Using equation (\ref{eq:T_ex}) one can switch from integration
over $d\tau_\nu$ to integration over $y\equiv h\nu/k_B T$ in the
range $y_0<y<\infty$, where $y_0\equiv h\nu/k_B T_{ex}$. The upper 
limit $h\nu/k_B T_{sh}$ can be extended to $\infty$ since most
of the exposed layer emissivity originates at temperatures 
$\sim T_{ex}$ corresponding to $y\sim y_0$.  
As a result, one obtains
\ba
&& \frac{dF_\nu^{ex}(a)}{dS}=2\pi
\left\{1+\exp\left[-\frac{\Sigma_0(a)\kappa_\nu}{\cos i}\right]\right\}
A_\nu(T_{ex}),
\label{eq:SED_outer1}
\ea
where
\ba
 A_\nu(T_{ex})& = & \alpha\frac{\kappa_\nu}{\kappa_P(T_\star)}
\int\limits_0^{T_{ex}} \frac{dT}{T}B_\nu(T)\psi(T)\nonumber\\
 & = & \alpha
\frac{2 h\nu^3}{c^2}\frac{\kappa_\nu}{\kappa_P(T_\star)}
\int\limits_{y_0}^\infty \frac{dy}{y}\frac{\psi\left(h\nu/ky\right)}{e^y-1},
\label{eq:A_nu}
\ea
is the emissivity of the exposed layer. Function $\psi(T)$ is defined as 
\ba
\psi(T)& \equiv & 4+\frac{d\ln\kappa_P(T)}{d\ln T}
\nonumber\\
& = & \frac{\int_0^\infty B_\nu(T)\kappa_\nu
\left(4+d\ln\kappa_\nu/d\ln\nu\right)d\nu}
{\int_0^\infty B_\nu(T)\kappa_\nu d\nu}.
\label{eq:psi}
\ea
Equations (\ref{eq:SED_outer1})-(\ref{eq:psi}) provide us 
with the method of calculation of the SED of the exposed layer 
that does not require an explicit solution of (\ref{eq:tem_calvet}) 
to obtain $T(\tau_\star)$ but at the same time fully accounts 
for the non-isothermal nature of the exposed layer.

\begin{figure}
\plotone{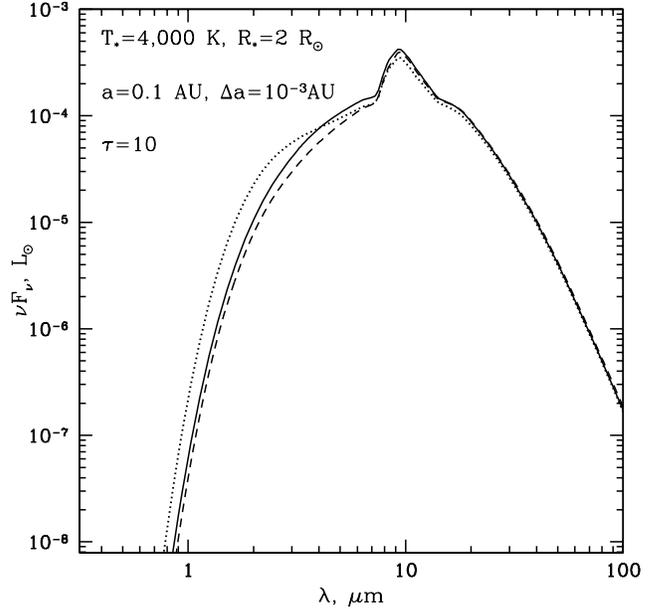}
\caption{
Same as Figure \ref{fig:AeBe_1} but for an annulus around 
T Tauri star ($T_\star=4000$ K, $R_\star=2~R_\odot$, 
$M_\star=0.5~R_\odot$). Annulus is placed at $a=0.1$ AU and
has width $\Delta a=0.001$ AU and vertical optical depth 
$\tau=10$ at $550$ nm.  
\label{fig:TT}}
\end{figure}

Full emissivity is given by the sum of $dF_\nu^{sh}(a)/dS$ 
and $dF_\nu^{ex}(a)/dS$.
In Figures \ref{fig:AeBe_1}-\ref{fig:TT} we display the 
SED produced by the thin disk annulus 
placed at different radii around different stars
(their parameters can be found in figure captions), 
assuming a fixed value of the flaring angle 
$\alpha=0.05$ and inclination $i=0^\circ$ for all of them. This choice
allows us to directly check our results against the calculations of 
DN03 who have made an exhaustive comparison 
between the two-temperature model 
and the full numerical treatment of the radiative 
and hydrostatic equilibrium of the disk, using both 
frequency- and angle-dependent radiation transfer in 
$z$-direction. 

Figures \ref{fig:AeBe_1}-\ref{fig:TT} show that at
short wavelengths our treatment (\ref{eq:SED_outer1})-(\ref{eq:psi}) 
of the temperature structure of the exposed layer reproduces 
full numerical SED remarkably well, much better than the two-temperature 
approximation. This is not
surprising since unlike our equation (\ref{eq:T_ex}) the 
two-temperature model does not capture the smooth 
transition between $T_{ex}$ and $T_{sh}$ in the exposed layer 
and this affects its performance at high frequencies. 

At longer wavelengths situation is more complicated. 
In particular cases represented in Figures \ref{fig:AeBe_1} 
and \ref{fig:TT} there is very nice agreement 
between our SED calculations and the numerical 
results of DN03 in the whole range of wavelengths. However, in the
case displayed in Figure \ref{fig:AeBe_50} situation is
not so simple: both our approach and the two-temperature 
approximation give very similar results but deviate significantly 
from the SED computed by DN03 at $\lambda\sim 100$ $\mu$m. 
This has to do with the fact that at longer wavelengths
disk spectrum becomes more sensitive to 
the details of the thermal structure of the shielded layer
which is treated similarly by both our method 
and the two-temperature approach of CG97. However, it has been 
previously shown by Dullemond, van Zadelhoff \& Natta 
(2002) that  in some cases neither equation (\ref{eq:tem_calvet}) 
nor (\ref{eq:T_CG}) reproduce the temperature structure of 
the shielded layer accurately enough, and this causes  
discrepancies at long wavelengths in spectral calculations. 
Nevertheless, the much better agreement at short wavelengths 
obtained with our procedure at the expense of only slightly 
increased (compared to the two-temperature model) 
complexity justifies its practical use in various applications.


\section{Discussion and Summary.}
\label{sect:discussion}


We have explored structure of the irradiated circumstellar disk 
based on a realistic temperature profile (\ref{eq:tem_calvet}), 
which is more accurate than the two-temperature approximation of CG97. Our results
on the disk SEDs presented in \S  \ref{sect:spectra} persuasively
demonstrate that this approach is superior to that of CG97
or DDN as it naturally reproduces the spectral behavior at 
short wavelengths without making any artificial assumptions.

This improvement is significant because of the recently
emerged interest to the structure of the SED around 
$\lambda\sim 2$ $\mu$m. Many circumstellar disks exhibit 
excess emission in this band (Hillenbrand \etal 1992) 
which has been generally interpreted as 
the evidence for the existence of the dust sublimation 
region in the inner parts of these disks (DDN). Examination of Figures
\ref{fig:AeBe_1} \& \ref{fig:TT} reveals that the use of the 
two-temperature approximation can significantly overestimate
the contribution of the disk to the total flux in this band.
As a result, one would {\it underestimate} the amount of
emission produced by the dust sublimation region and this 
can seriously affect the interpretation of the data. 
On the contrary, our approach
reproduces realistic disk SEDs (obtained by DN03 using rather 
time-consuming procedure for solving full radiation transfer)
very accurately at short wavelengths and, thus, can be 
relied upon when determining the emissivity of the 
dust sublimation region from the data. Note that our method
of SED calculation based on equation (\ref{eq:tem_calvet})
is no more computationally intensive than the two-temperature 
approximation of CG97 which is routinely used for
analyzing  protoplanetary disk spectra.

Another goal of this study was to introduce and test the 
constant $g_z$ approximation (\S \ref{subsect:g_z}) 
for treatment of the irradiated circumstellar disk structure. 
We have demonstrated by comparison with numerical results 
(\ref{sect:pure_radial}) that within its region of applicability
[see equation (\ref{eq:validity})] this approximation
works very well. Its use has allowed us to obtain analytical 
solutions for the vertical disk structure and to determine 
the position of the disk surface as a function of distance 
from the star rather accurately through the disk and stellar 
parameters. Improved description of the density and 
temperature structure of gas in the disk photosphere 
at $\tau_\star\sim \tau_c$ obtained in \S \ref{subsect:g_z} 
allows one to improve the predictions of molecular line 
intensities compared to the two-temperature approximation 
[see Dullemond \etal (2002)
for comparison of the line intensities obtained using full
radiation transfer and its simplified representation given 
by equation (\ref{eq:tem_calvet})].

Finally, solutions presented in \S \ref{subsect:g_z} 
allow fast and efficient exploration of the large phase 
space of circumstellar disk parameters 
when fitting disk SEDs. Their analytical 
nature will also be useful for  treating more 
complicated problems, e.g. determining the structure 
of the dust sublimation region, ice lines, and so on.\\

\acknowledgements 

RRR thankfully acknowledges the financial support by the 
Canada Research Chairs program, W. M. Keck Foundation, and 
NSF via grant PHY-0070928. FDC acknowledges financial 
support through the fellowship of the DGEP-UNAM, and 
from the EU Community under the Marie Curie Research Training 
Network ''JETSET''. FDC appreciates the hospitality and 
financial support of the Institute for Advanced Study (IAS) 
during the visit when this work has been initiated. 

\appendix


\section{$\psi$-factors for optically thin disks.}
\label{ap:psi_factors}


Distant parts of the circumstellar disk where 
$\Sigma_0$ is small and temperature is low can be optically thin
to the radiation of the shielded layer ($\Sigma_0\kappa_P(T_{sh})
\lesssim 1$) and/or the exposed layer ($\Sigma_0\kappa_P(T_{ex})
\lesssim 1$). To account for this possibility one introduces factor 
$\psi_{sh}<1$ defined as the ratio of the shielded layer emissivity
to its value in the optically thick case. 
Factor $\psi_{ex}<1$ is defined as the fraction
of radiation emitted by the exposed layer that gets absorbed by 
the shielded layer. 

There are different ways of defining $\psi$-factors. 
In particular, Chiang \etal (2001) used simply 
\ba
\psi_{sh}^{\rm C}(T_{sh})=
1-\exp\left[-\Sigma_0\kappa_P(T_{sh})\right],~~~
\psi_{ex}^{\rm C}(T_{ex})=1-\exp\left[-\Sigma_0\kappa_P(T_{ex})\right],
\label{eq:psi_irr_CG}
\ea
while DDN suggested more accurate expressions for $\psi_{sh}$ and $\psi_{ex}$:
\ba
&&\psi_{sh}^{\rm DDN}(T_{sh})=\frac{\int_0^\infty B_\nu(T_{sh})
\left[1-\exp\left(-\Sigma_0\kappa_\nu\right)\right] d\nu}{\sigma T_{sh}^4},
\label{eq:psi_irr_DDN}\\
&&\psi_{ex}^{\rm DDN}(T_{ex})=\frac{\int_0^\infty B_\nu(T_{ex})\kappa_\nu
\left[1-\exp\left(-\Sigma_0\kappa_\nu\right)\right] d\nu}{\kappa_P(T_{ex})}.
\label{eq:psi_ex_DDN}
\ea

All of these expressions assume that corresponding layers of the
disk emit pure blackbody radiation which is a reasonable assumption
for the isothermal shielded layer (and so $\psi_{sh}^{\rm DDN}$ should
be accurate). However, the temperature of the exposed layer varies 
quite dramatically when $1\lesssim \tau_\star\lesssim \tau_c$ and its
spectrum must deviate from $B_\nu(T_{ex})$.
For this reason, for $\psi_{ex}$ we use the following 
expression:
\ba
\psi_{ex}(T_{ex})=\frac{\int_0^\infty A_\nu(T_{ex})
\left[1-\exp\left(-\Sigma_0\kappa_\nu\right)\right] d\nu}
{\int_0^\infty A_\nu(T_{ex})d\nu},
\label{eq:psi_ex_RD}
\ea
where $A_\nu(T)$ is the true emissivity of the 
exposed layer given by equation (\ref{eq:A_nu}).
Clearly, all $\psi$-factors reduce to unity in the optically thick disk 
($\Sigma_0\kappa_P(T_{sh}),~\Sigma_0\kappa_P(T_{ex})\gtrsim 1$).

\begin{figure}
\plotone{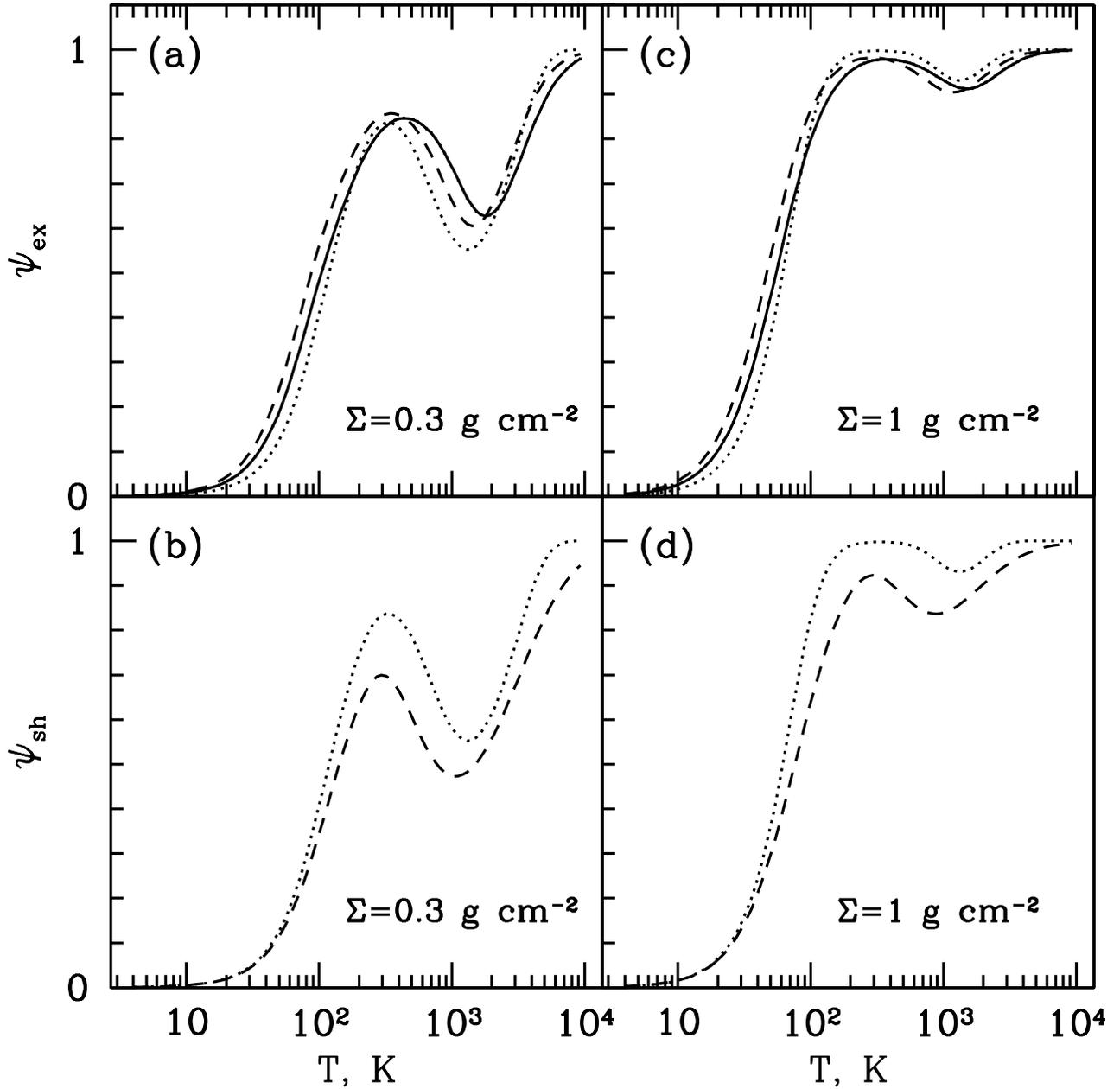}
\caption{
(a,c) Plots of $\psi_{ex}$ ({\it solid curve}), $\psi_{ex}^{\rm C}$ 
({\it dotted curve}), and $\psi_{ex}^{\rm DDN}$ ({\it dashed curve})
for two different values of disk surface density: $\Sigma=0.3$ g 
cm$^{-2}$ (a) anf $\Sigma_0=1.0$ g cm$^{-2}$ (c). (b,d) 
The same for $\psi_{sh}^{\rm C}$ 
({\it dotted curve}), and $\psi_{sh}^{\rm DDN}$ ({\it dashed curve}).
\label{fig:psi}}
\end{figure}

In Figure \ref{fig:psi} we plot different $\psi$-factors for two
values of $\Sigma$. One can see that all three prescriptions 
(\ref{eq:psi_irr_CG}), (\ref{eq:psi_ex_DDN}), and 
(\ref{eq:psi_ex_RD}) for $\psi_{ex}$ give rather similar 
results. At the same time, simple expression (\ref{eq:psi_irr_CG})
for $\psi_{sh}$ advocated by Chiang \etal (2001) differs
from the more accurate prescription (\ref{eq:psi_irr_DDN}) 
at the level of tens of per cent in some temperature ranges.
In this work we use $\psi_{sh}$
and $\psi_{ex}$ given by equations (\ref{eq:psi_irr_DDN}) and 
(\ref{eq:psi_ex_RD}) correspondingly. 


\section{Radial properties of irradiated disks.}
\label{ap:ap1}


In the case of lamp-post illumination one immediately finds 
from (\ref{eq:lamp_post}) and (\ref{eq:T_irr}) that
\ba
&& T_{sh}(a)=T_\star\left(\frac{\phi\eta}{2}\right)^{1/4}
\left(\frac{a}{R_\star}\right)^{-3/4},~~~h_{sh}(a)=h_\star
\left(\frac{\phi\eta}{2}\right)^{1/8}\left(\frac{a}{R_\star}
\right)^{9/8},
\label{eq:lp_case}
\ea
where $h_\star\equiv\Omega_\star^{-1}(kT_\star/\mu)^{1/2},~
\Omega_\star\equiv(GM_\star/R_\star^3)^{1/2}$. 
Value of $\Lambda$ in this case is given by
\ba
\Lambda_{lp}\approx\frac{\kappa(T_\star)\Sigma_\star}
{(2\pi)^{1/2}\eta\lambda\tau_c}
\left(\frac{a}{R_\star}\right)^{1-\delta},
\label{eq:Lambda_lp}
\ea

In the case of point-like illumination the substitution of  
equation (\ref{eq:point_like}) with $H_1\approx H_0=\lambda h_{sh}$
into definition (\ref{eq:T_irr}) yields a differential 
equation for $T_{sh}$, which can be easily
solved if $\lambda$ is roughly constant. One finds
\ba
&& T_{sh}(a)=T_\star\left(\frac{\phi\lambda}{7}\frac{h_\star}{R_\star}\right)^{2/7}
\left(\frac{a}{R_\star}\right)^{-3/7},~~~h_{sh}(a)=h_\star
\left(\frac{\phi\lambda}{7}\frac{h_\star}{R_\star}\right)^{1/7}
\left(\frac{a}{R_\star}\right)^{9/7},\nonumber\\
&& \alpha=2\left(\frac{\phi\lambda}{7}\frac{h_\star}{R_\star}\right)^{8/7}
\left(\frac{a}{R_\star}\right)^{2/7}.
\label{eq:ps_case}
\ea
For point-source illumination $\Lambda$  is given by 
\ba
\Lambda_{ps}\approx\frac{\kappa(T_\star)\Sigma_\star}
{(8\pi)^{1/2}\lambda\tau_c}
\left(\frac{\phi\lambda}{7}\frac{h_\star}{R_\star}\right)^{-8/7}
\left(\frac{a}{R_\star}\right)^{-(\delta+2/7)}.
\label{eq:Lambda_ps}
\ea
Equations (\ref{eq:Lambda_lp}) and (\ref{eq:Lambda_ps}) demonstrate that
$\lambda=(2\ln\Lambda)^{1/2}$ is indeed a very weak function 
of $a$ when $\kappa(T_\star)\Sigma_0(a)\gg 1$, thus justifying the 
constancy of the ratio $H_1/h_{sh}$ in the inner optically thick 
parts of the disk.


\section{Calculation of the radial structure.}
\label{ap:ap3}


Numerical solutions for the disk structure based on 
approximation (\ref{eq:tem_calvet}) are obtained using
iterative procedure. On the first step we adopt some 
more or less arbitrary initial radial profile for 
$H_1(a)$, which yields $\alpha(a)$ and $T_{sh}(a)$ through 
equations (\ref{eq:alpha}) and (\ref{eq:T_irr}).
This allows us to integrate equations 
(\ref{eq:hor_vert_opt_depth}) and (\ref{eq:hydro}) using
temperature profile (\ref{eq:tem_calvet}) and the ideal 
gas law, resulting in distributions of $\tau_\star(a,z)$, 
$\rho(a,z)$, and $P(a,z)$. With this we can determine 
new profile of $H_1(a)$ as the vertical height where 
$\tau_\star=1$ at every $a$. Then we repeat iteration
until the convergence is obtained.

This procedure is generally quite unstable because of the
presence of derivative of $H_1/a$ with respect to $a$ needed 
for calculation of $\alpha$. To abate this problem we
average each value of $H_1(a)$ over the $5$
neighbouring grid points. Also,  we put a ``limiter''
in the determination of $\alpha$, i.e. we limit the
relative change of $\alpha$ with respect to its value
in the previous iteration to be no more that 10\%. 
These simple steps allow us to eliminate spurious 
oscillations arising in the determination of $\alpha$.
We found this procedure to more robust and accurate than the one
proposed by Chiang et al. (2001) as it allows us to significantly 
increase radial resolution and obtain very accurate results for the
radial disk structure without giving
rise to parasitic instabilities.


\begin{center}
\begin{deluxetable}{ l l l l l l l l }
\tablewidth{0pc}
\tablecaption{Representative models of circumstellar disks. 
\label{table}}
\tablehead{
\colhead{Model}&
\colhead{$M_\star$ $[M_\odot]$}&
\colhead{$R_\star$ $[R_\odot]$}&
\colhead{$T_\star$ [K]}&
\colhead{$R_{in}$ [AU]}&
\colhead{$R_{out}$ [AU]}&
\colhead{$\Sigma_1$ [g cm$^{-2}$]}&
\colhead{Object}
}
\startdata
I & $2$ & $2$ & 10000 & 1 & 300 & 1000 & Herbig Ae/Be \\
II & $0.5$ & $2$ & 4000 & 0.1 & 300 & 100 & T Tauri \\
III & $0.1$ & $1.3$ & 2600 & 0.033 & 30 & 100 & Brown Dwarf \\
\enddata
\end{deluxetable}
\end{center}

\end{document}